\documentclass{svproc}

\usepackage[utf8]{inputenc}
\usepackage[T1]{fontenc}
\usepackage[linesnumbered,ruled,vlined]{algorithm2e}
\usepackage{algpseudocode}
\usepackage{hyperref}
\usepackage{booktabs}
\hypersetup{colorlinks=true,linkcolor=black,citecolor=black,filecolor=magenta,urlcolor=cyan}
\usepackage{amsmath, graphicx}
\usepackage[shortlabels]{enumitem}

\usepackage{tikz}
\usetikzlibrary{arrows}
\usetikzlibrary{shapes}

\usetikzlibrary{positioning, fit, calc}   
\tikzset{block/.style={draw, thick, text width=2cm ,minimum height=1.3cm, align=center},   
line/.style={-latex}     
}

\algdef{SE}[VARIABLES]{Variables}{EndVariables}
  {\algorithmicvariables}
  {\algorithmicend\ \algorithmicvariables}
\algnewcommand{\algorithmicvariables}{\textbf{global variables}}

\algdef{SE}{Function}{EndFunction}[1]{\textbf{function}\(\mbox{#1}\)}{\textbf{end function}}%

\begin{document}




\title{Committee selection in DAG distributed ledgers and applications}

 \author{Bartosz Ku\'{s}mierz$^{1,2}$, Sebastian M\"uller$^{3}$, Angelo Capossele$^1$ }

\authorrunning{Bartosz Ku\'{s}mierz, et al.}

\institute{$^1$IOTA Foundation, 10405 Berlin, Germany \\
$^2$Department of Theoretical Physics,\\ Wroclaw University of Science and Technology, Poland\\
$^3$Aix Marseille Universit\'e, CNRS, Centrale Marseille, I2M - UMR 7373, 13453 Marseille, France}





\maketitle

\begin{abstract}
In this paper, we propose several solutions to the committee selection problem among participants of a DAG distributed ledger. Our methods are based on a ledger intrinsic reputation model that serves as a selection criterion. The main difficulty arises from the fact that the DAG ledger is a priori not totally ordered and that the participants need to reach a consensus on participants' reputation.

Furthermore, we outline applications of the proposed protocols, including:  {\it (i)} self-contained decentralized random number beacon; {\it (ii)} selection of oracles in smart contracts; {\it (iii)}  applications in consensus protocols and sharding solutions. 

We conclude with a discussion on the security and liveness of the proposed protocols by modeling reputation with a Zipf law. 
\end{abstract}

%




\section{Introduction}


In distributed ledger technologies (DLTs), committees play essential roles in various applications, e.g.,  distributed random number generators, smart contract oracles, consensus mechanisms, or scaling solutions.

The most famous example of permissionless DLT is the Bitcoin blockchain introduced in the whitepaper \cite{bitcoin} by Satoshi Nakomoto.  The blockchain enables network participants to reach consensus in a trustless peer-to-peer network using the so-called proof of work (PoW) consensus protocol. In PoW based blockchains, participants need to solve a cryptographic puzzle to issue the next block. The higher the computing power of a participant, the higher the chances to produce the next block. 

In the past years, another consensus mechanism, called Proof-of-Stake (PoS), became popular.
In contrast to PoW, in PoS-based cryptocurrencies, the next block's creator is chosen depending on its wealth or stake \cite{PopovNxt} and not on its computing power.

 
DLTs already found multiple applications in the financial sector, including online value transfer, digital assets management, data marketplace \cite{gudici2020a,zheng2016}. Successive projects, inspired by Bitcoin, started an entirely new field of smart contracts, which are used for settling online agreements without the intermediary and the open possibility of decentralized autonomous organizations \cite{daooo1,Peters2016}. 

Nevertheless, blockchain-based DLTs have problems, which become apparent when the network participants start using full block capacity. As the number of transactions issued by the users became significantly bigger than what can fit into the block, fees began to increase considerably. Increasing the throughput is one of the main motivations behind the DLTs' scaling problem, which leads to the (in-)famous blockchain trilemma \cite{tricurse}. The blockchain trilemma states that DLT can have up to two out of three desired properties: scalability, security, decentralization. 
The most problematic part of the trilemma for blockchains is scalability. 

Proposed solutions that aim at increasing DLTs scalability and flexibility include increasing block size and issuance frequency, sidechains \cite{croman2016}, ``layer-two'' solutions like Lightning Network \cite{light}, different consensus mechanisms \cite{casper},  sharding \cite{luu2016}. A different approach is to change the underlying data structure from a chain to a more general directed acyclic graph (DAG) \cite{I_WP,I_EQ}.  

DAGs have been adopted by a variety of DLT projects including  IOTA \cite{I_WP}, Obyte 
\cite{byteball}, SPECTRE \cite{spectre}, Nano \cite{nano}, Aleph Zero \cite{alephzero}. While those projects utilize the DAG structure differently and adapted different consensus mechanisms, in most of them, transactions are graph vertexes that reference multiple previously issued transactions. This property assures that the graph is acyclic, and the data structure grows with time. 

\subsection{DAG based cryptocurrencies}
An example of a project that utilizes DAGs is IOTA as described in the original whitepaper \cite{I_WP}, which requires every transaction in the DAG, also called the {\it Tangle}, to reference exactly two other transactions directly. Transactions also indirectly reference other transactions, we say that $y$ indirectly approves $x$ if there is a directed path of references from $y$ to $x$. As the ledger grows, each accepted transaction gains indirect references, which are assured by the default tip selection mechanism \cite{I_WP,I_EQ,pcdet,iotaproperties}. The set of indirect references of a given transaction is interpreted as the number of confirming transactions and play the same role as the confirming blocks in Bitcoin, i.e., the more indirect references a transaction gains, the more likely it is to remain a part of the ledger.

The currently deployed implementation of the IOTA  is a form of technology prototype \cite{githubiota}, hereafter referred to as Coordinator-based-IOTA, and differs from the original whitepaper version. The most crucial difference is that the consensus is based on the so-called {\it milestones} -  special transactions issued periodically by a privileged node called {\it Coordinator}.  Every transaction referenced (directly or indirectly) by milestone is considered confirmed. While such a system is centralized, authors in \cite{coordicide} propose a decentralized solution dubbed {\it Coordicide}, referred to as post-Coordicide IOTA.  
One key element in Coordicide is to replace the Coordinator with the decentralized consensus protocol called Fast Probabilistic Consensus (FPC).  FPC is a scalable Byzantine resistant voting scheme \cite{fpcsim,manaFPC,FPC} where nodes vote in rounds. In each round, participants query a random subset of online nodes in the network. Opinions in the next round depends on the received queries and a random threshold.
When most of the nodes in the network 
use the same random threshold the system reaches unanimity. The common random thresholds enable the protocol to break meta-stable situations \cite{Po:video}.



\subsection{Contributions}

Let us define a committee as a group of usually trustworthy nodes, selected to execute a special task. We assume that participants take part in a DLT that supports a reputation system. 
Every node can issue transactions, which from now on we call {\it messages}. We use the name message to indicate that those objects can include generic data and not only token transfers. The reputation system is needed as a criterion to select a subgroup of all participants and to mitigate Sybil attacks, a common threat in permissionless systems. 

We concentrate on the more difficult case of DAG-based DLTs, which promise better scalability and decentralization than blockchains. Unlike blockchains, which are totally ordered by nature, DAGs lack natural ``reference points'' that could be used to determine the reputation of the participants. The confined structure of blockchains allows for using reputation calculated for a specific block defined on the protocol level. The complex structure of DAGs does not allow for the adaptation of an analogical rule.

This paper proposes a series of committee selection mechanisms for DAG-based DLTs with reputation system. The committee selection process runs periodically and depends on the reputation of nodes, which is stake or delegated stake. 
More specifically, the contributions of this paper are the following:
\begin{enumerate}
    \item we propose several protocols to select a committee in permissionless decentralized systems\label{commmmm};
    \item we analyze the token distribution of a series of cryptocurrency projects;
    \item we model the reputation distribution and analyze the security of the proposed protocols.
    \item we discuss several applications of committee selection protocols, including decentralized random number beacon, smart contracts, consensus mechanism, and sharding solutions.
\end{enumerate}

\subsection{Outline}

The article is organized as follows. In the next section, we discuss previous works related to this paper, mainly related to different kinds of decentralized random number beacon and reputation systems. In section \ref{sec:prot}, we specify the assumptions on the DLT that are required for our protocol to work. Section \ref{sec:com} is devoted to the committee selection, where we give three different methods of achieving consensus on the reputation values. In section \ref{sec:secur}, we discuss the security of our proposal by modeling reputation distribution with a Zipf law. Applications of our protocols to dRNG, smart contracts, consensus mechanism, and scaling are discussed in section \ref{sec:app}. Finally, section \ref{sec:dis}  outlines  further  research  direction  and  concludes  the paper.

\section{Related work}

\subsection{Reputation and committees in DLT}

A reputation  system  in  a    DLT is any mapping that assigns real numbers to the network participants. Reputation can be objective when all of the participants agree on the exact values of the reputation, or subjective when different nodes have different perception of the reputation. However, for the subjective reputation to maintain its utility and play the same function as in social systems, network users should have at least an approximate consensus on its values. 

All PoS consensus mechanisms induce a reputation system where the user's reputation equals staked tokens \cite{casper}. In the same way,  delegated PoS (DPoS) protocols, where staked tokens are delegated to other nodes \cite{Tron,EOS}, define a natural reputation system. Highest DPoS nodes form a committee of block validators and produce the next blocks. 
The consensus among the fixed-size committee is easier to achieve than in open systems. An interesting variation on DPoS systems is {\it mana} introduced in the post-Coordicide IOTA network \cite{coordicide}. This reputation system takes advantage of each issued message, which temporally grants mana to a certain node. 

Multiple implementations of sharding in blockchains also require a random assignment of the block validators, e.g., \cite{scal1}. If validators can not predict which shard they will be assigned, then any collusion is significantly hampered. When the network also uses a reputation system, the sharding process can be improved by assigning approximately the same reputation into each shard. An example of such protocol is RepChain \cite{repchain}, which assigns reputation to nodes based on their behavior in the previous rounds. 

\subsection{Decentralized random numbers generation}

Both randomness and reputation systems can improve the security, scalability, and liveness of the DLTs. A random number beacon, as introduced by Rabin \cite{RABIN1983256}, is a service that broadcasts a random number at regular intervals. Randomness produced by an ideal beacon cannot be predicted before being published; however, this assumption is hard to achieve in centralized systems. 

The development of decentralized random number generators (dRNG) tries to address those problems. In general, a dRNG should provide unpredictable and unbiased randomness, which can not be controlled nor easily biased by a single malicious actor. There are different proposals for dRNGs in the literature.

Authors in \cite{blockvdf} discuss the extraction of randomness from public blockchains using hashes of blocks. However, certain concerns regarding those solutions have been raised in \cite{btcr,Popov2017}. Other proposals like RandHound and RandHerd utilize publicly verifiable secret sharing schemes \cite{randHH} to generate a collective key shared between committee nodes. If more than a certain threshold of partial secrets are published, the network can recover the random number, i.e., $(t,n)$-threshold security model. Other proposals include smart contracts, e.g., RANDAO used in Ethereum \cite{randao1}. Security in such solutions is achieved with the risk of fund confiscation.

An interesting research direction that can improve security and prevent manipulation of the generated randomness are verifiable delay functions (VDFs) \cite{blockvdf}. VDFs take a fixed amount of time to compute, can not be parallelized but can be verified quickly  \cite{wes,pietrz}. However, VDF calculations are costly.  Moreover, this approach requires further research to ensure honest users have access to the fastest application-specific integrated circuits (ASICs) specialized in a given VDF.

\section{Protocol overview and setup} \label{sec:prot}

We propose a protocol for committee selection in DAG-based DLTs with a subjective reputation system. Every node has a reputation, but a priori, there is no perfect consensus on the values of the reputation. We assume that each vertex of a DAG determines the view of the reputation, and subjectivity comes from the fact that there is no unique way of choosing the ``reference vertex''. For example, if nodes adopt the simple rule {\it reputation should be calculated based on the most recent received message}, then due to network delay, two different nodes disagree on which is the most recent.

The committee selection is the process of appointing nodes with a sufficiently high reputation. 

To perform the three phases above, we require the underlying DAG to verify the following properties.

\begin{enumerate}
    \item[P1] \label{itm:1b} The DAG grows in time, and incoming messages are new vertexes of the DAG.

    \item[P2] The DAG allows for the message exchange of {\it application messages} (optional). 
    
    \item[P3] The DAG is immutable and provides a strict criterion for an approximate time of message creation.  
    \label{immut}
    \item[P4] The subjective reputation of the nodes can be read from DAG and is determined by the vertex (different nodes can read reputation from different vertexes). \label{ppp4} 
    
    \item[P5] \label{itm:5b} Messages in the DAG are signed by the nodes and can not be counterfeited (i.e., a malicious node can not fake origin of the message).

\end{enumerate}
Immutability, P\ref{immut}, guarantees that no message can be removed from the ledger nor new messages can be added in a part of the ledger that suggests it was issued in the past (by attaching it deep into the DAG). Property P\ref{immut} can be achieved using any of the following methods:

    \begin{enumerate}[(i)]
        \item Messages are equipped with  {\it enforceable} timestamps. Messages with wrong  timestamps are rejected by the network. \label{timest} 
        
        \item Special {\it partial order generating} messages are periodically issued into the DAG.   \label{order} 
    \end{enumerate}

Enforceable timestamps mentioned in \ref{timest} can be achieved when honest nodes automatically reject messages with timestamps too far in the past or the future. A certain level of desynchronization must be allowed to account for the network delay and differences in local clocks. However, we require that the network has a specific bound above which no message with lower timestamp will be accepted into the ledger (even accounting for network delays). In the edge cases of timestamps, when part of the network thinks that timestamp is valid and other does not, it is necessary to run a certain kind of consensus mechanism.

Partial ordering generating (POG) messages, \ref{order},  can be a result of consensus among the nodes, e.g., nodes vote on them. Other options are {\it proof-of-authority} type of consensus where privileged ``validator'' nodes issue those POG messages. An example of this consensus type is the Coordinator-based IOTA \cite{githubiota}, where a particular entity called {\it Coordinator} issues milestones. Similarly, Obyte uses main chain transactions (MCT), which are indicated by trusted {\it witnesses} \cite{byteball}. Note that when a node issues a message after the $i$th milestone/MCT, it can be approved only by a $(i+1)$th order generating message. This procedure generates a certain kind of logical timestamps provided by the milestones/MCT.

There are two natural ways of determining the reputation value: 
    \begin{enumerate}[(i)]
        
        \item the reputation is summed over all of the messages approved by a given message; \label{pastcone}       
        
        \item if timestamps are available, the reputation can be summed over all of the messages with timestamps smaller than DAG vertex's timestamp. 
    \end{enumerate}

An example of the DAG's reputation calculated using the method in \ref{pastcone}  is presented in Fig.\ \ref{pastconefig}.

\begin{figure}
\begin{center}
\resizebox{0.4\textwidth}{!}{
\begin{tikzpicture}
\tikzset{
node_style/.style={circle,draw=black,fill=black!10!, minimum size=0pt},
node_styleblue/.style={circle,draw=black,fill=blue!30!, minimum size=0pt},
node_stylegreen/.style={circle,draw=black,fill=green!30!, minimum size=0pt},
node_styledarkblue/.style={circle,draw=black,fill=blue!50!, minimum size=0pt},
node_styledarkgreen/.style={circle,draw=black,fill=green!50!, minimum size=0pt},
node_stylethick/.style={circle,draw=black,fill=black, minimum size=0pt},
edge_style/.style={draw=black,  thin, ->},
edge_stylethick/.style={draw=black,  ultra thick, ->},
label/.style={sloped,above},
 sh/.style={ shade, shading=axis, left color=green!30, right color=blue!30,
    shading angle=90 } 
}

\node[draw=red, ellipse,minimum height=1.5cm,minimum width=2.5cm,fill=red!30!] (genesis) at (0,-1) {Genesis};

\node[draw, circle, align=center,node_stylegreen] (N1) at (-3,2) {Node B \\ $+10$};
\node[draw, circle, align=center,sh] (N2) at (-1,1.5) {Node C \\ $+10$};
\node[draw, circle, align=center,node_styleblue] (N3) at (1,2) {Node A \\ $+10$};
\node[draw, circle, align=center,node_styleblue] (N4) at (3,2) {Node B \\ $+5$};

\draw [edge_stylethick]  (N1) edge  (genesis);
\draw [edge_stylethick]  (N2) edge  (genesis);
\draw [edge_stylethick]  (N3) edge  (genesis);
\draw [edge_stylethick]  (N4) edge  (genesis);

\node[draw, circle, align=center,node_stylegreen] (N5) at (-2,5) {Node B \\ $+15$};
\node[draw, circle, align=center,sh] (N6) at (0,3.7) {Node C \\ $+5$};
\node[draw, circle, align=center,node_styleblue] (N7) at (2,5) {Node A \\ $+5$};

\draw [edge_stylethick]  (N5) edge  (N1);
\draw [edge_stylethick]  (N5) edge  (N2);

\draw [edge_stylethick]  (N6) edge  (N2);
\draw [edge_stylethick]  (N6) edge  (genesis);

\draw [edge_stylethick]  (N7) edge  (N3);
\draw [edge_stylethick]  (N7) edge  (N4);

\node[draw, circle, align=center,node_styledarkgreen] (N8) at (-1.8,8) {Node A \\ $+5$};
\draw [edge_stylethick]  (N8) edge  (N6);
\draw [edge_stylethick]  (N8) edge  (N5);

\node[draw, circle, align=center,node_styledarkblue] (N9) at (2.2,8) {Node C \\ $+5$};
\draw [edge_stylethick]  (N9) edge  (N6);
\draw [edge_stylethick]  (N9) edge  (N7);

\node[draw, circle, align=center,node_style] (N10) at (0.2,7) {Node A \\ $+25$};
\draw [edge_stylethick]  (N10) edge  (N5);
\draw [edge_stylethick]  (N10) edge  (N7);

\end{tikzpicture}

}
\caption{Each message grants a particular reputation to one node. The reputations are summed over all of the messages approved by a given message.  The reputation calculated for the dark blue message takes into account all blue messages:  node A: 15, node B:5, node C: 20.  The reputation for the dark green message is obtained by summing over all green messages:  node A: 5, node B:25, node C 15.  } \label{pastconefig}
\end{center}
\end{figure}


\section{Committee selection}\label{sec:com} 

When all users have the same view on the reputation,  a natural choice for the committee is to take the top $n$ reputation nodes. 

An alternative option is to perform a lottery with a probability dependent on the reputation of nodes. The list of lottery participants consists of $k>n$ nodes with the highest reputation. Using the last random number $X$ produced by the previous committee, nodes calculate the coefficients:
\[
q_i=f(X, id_i) \cdot \frac{r_i}{\sum_{j=1}^{k} r_j},
\]
where $f$ is a cryptographic hash function with values in $[0,1]$, $i$ is the index of a node, $r_i$ its reputation, and $id_i$ is its identifier. Then, the committee members would be the $n$ nodes with the highest $q_i$.   

Cryptocurrency projects are known for their high concentration of  hashing power or (staked) tokens, and opening the possibility of low reputation nodes being members of the committee may lead to a decrease in security. Thus, we recommend selecting the top $n$ reputation nodes.  We discuss this issue in more detail in section \ref{sec:secur}.

In the following, we present three different methods to find consensus on the nodes' reputations and, therefore, on the members of the committee. We describe the methods for DAGs with timestamps; the corresponding versions for DAG with POG messages are straightforward adaptations. We assume that we want to select the committee at time $t_{C}$, and $D$ is the bound for accepting dRNG messages in the DAG, i.e., no honest node will accept dRNG messages with timestamps different by more than $D$ from its local time.

\subsection{Application message
}
The first method of committee selection requires all nodes interested in the committee participation to prepare a special {\it application} message. This message determines the value of the reputation of a given node by its timestamp. Application messages can be submitted within an appropriate time window.  

Let us denote the application time window length by $\Delta_A$, then the time window is 

\[
[t_{C}-D -\Delta_A, t_{C}-D].
\]

Application messages are used to extract {only} the reputation of the issuing node, i.e., the reputations of two potential committee members are deduced from different messages. If a node sends more than one application message, the other participants do not consider this node for the selection process. Since messages require node's signatures, this type of malicious behavior can not be imitated by an attacker. 

This method of committee selection has the advantage that only online nodes can apply. 
Moreover, if a particular node does not want to participate in the committee, e.g., due to planned maintenance, insufficient resources, it can decide not to apply.  In general, applications should not be mandatory as it is hard to enforce.

The committee selection procedure is open, and any node can issue an application message. However, the committee is formed from top reputation nodes, and low reputation nodes are unlikely to get a seat. Applications issued by low reputation nodes are likely to be not only redundant but possibly problematic when the network is close to congestion. 
Thus we propose the following improvements to the application process.

A node is said to be $M_\ell$ if, according to its view of reputations, it is among the top $\ell$ reputation nodes. Then nodes produce application messages according to the following:

If a node $x$ is $M_{2n}$ then it issues an application at the time $t_{C}-D -\Delta_A$.

For $k >2$ a node $x$ which is $M_{n\cdot k}$ but not $M_{n\cdot (k-1)}$ submits a committee application only if at the time $t_{C}-D -\Delta_A /({\ell-1})$ (according to $x$'s local time perception) there is less than $n$ valid application messages with stated reputation greater then the reputation of $x$. The timestamp of such application message should be $t_{C}-D -\Delta_A /({\ell-1})$. An example of pseudocode for scheduling the send of an application message is presented in algorithm \ref{a1}.

\SetKwInOut{Input}{input}
\SetKwInOut{Output}{output}
\SetKwInOut{Require}{require}

\begin{algorithm}[t]
\DontPrintSemicolon
\Require{
    nodes\_reputation(time) $\rightarrow$ {\it descending ordered list of nodes' reputation at input time;}\\
    get\_position(ID, ordered\_list\_of\_nodes) $\rightarrow$ {\it ID's position on the\\ given list;}\\
    get\_reputation(ID, time) $\rightarrow$ {\it ID's reputation at the given time;}\\
    application\_msg\_with\_rep\_higher\_than(reputation) $\rightarrow$ {\it amount\\ of application messages in DAG with reputation higher than\\ input;}
    }
\Input{
    $t_C$: {\it time of committee selection;}\\
    $D$: {\it bound for accepting dRNG messages in the DAG;}\\
    $\Delta_A$: {\it application time window length;}\\
    max\_k: {\it maximum index value for sending an application msg;}\\
    n: {\it size of the committee;}\\
    my\_ID: {\it ID of the node;}
    }
\If{(time = $t_C-D-\Delta_A$)}{
    k $\leftarrow$ 2\;
    \While{k $\leq$ max\_k}{
        wait\_until$(t_C-D-\Delta_A/(k-1))$\;
        rep\_list $\leftarrow$ nodes\_reputation($t_C-D-\Delta_A/(k-1)$)\;
        my\_ell $\leftarrow$ get\_position(my\_ID, rep\_list)\;
        my\_rep $\leftarrow$ get\_reputation(my\_ID, $t_C-D-\Delta_A/(k-1)$)\;
        \If{(application\_msg\_with\_rep\_higher\_than(my\_rep) $\geq$ n)}{
            \textbf{return}\;
        }
        \If{(my\_ell $\leq$ $k*n$)}{
            send\_application\_message()\;
            \textbf{return}\;
        }
        k $\leftarrow$ $k+1$\;
    }
}
\textbf{return}
\caption{Application message send scheduler. }
\label{a1}
\end{algorithm}

\subsection{Checkpoint selection}
We introduce checkpoints, similar to POG messages mentioned in 
\ref{order}.  The checkpoint is unpredictably obtained from ordinary messages, i.e., it is impossible to say in advance that any particular message will become a checkpoint at the time of issuance. To achieve that, we require a single random number $X$.
 
The reputation of the nodes is calculated for this checkpoint message. Moreover, checkpoints can suggest which nodes are online. For example, using the following rule: if the last message in a past cone of checkpoint issued by a node $x$ is older than a certain threshold, we consider this node offline. 

The random number $X$ can be the last random number produced by the previous committee, e.g., using the dRNG described in subsection \ref{sec:dRNG}, or can come from a different source. If the random number $X$ is revealed at time $t_{C}$, the time the committee selection should start, the checkpoint is the first message with a timestamp smaller than
\[
t_{C} - D - X.
\]
Possible ties are broken with lower hash.

The role of the random number $X$ is to improve security. If no participant, including a potential attacker, can predict the timestamp of the future checkpoint, successful attacks are hindered or made impossible.  In the following, we describe one attack scenario in the absence of the random factor $X$.

\begin{figure}[ht]
\begin{center}   
  \includegraphics[width=0.7\textwidth]{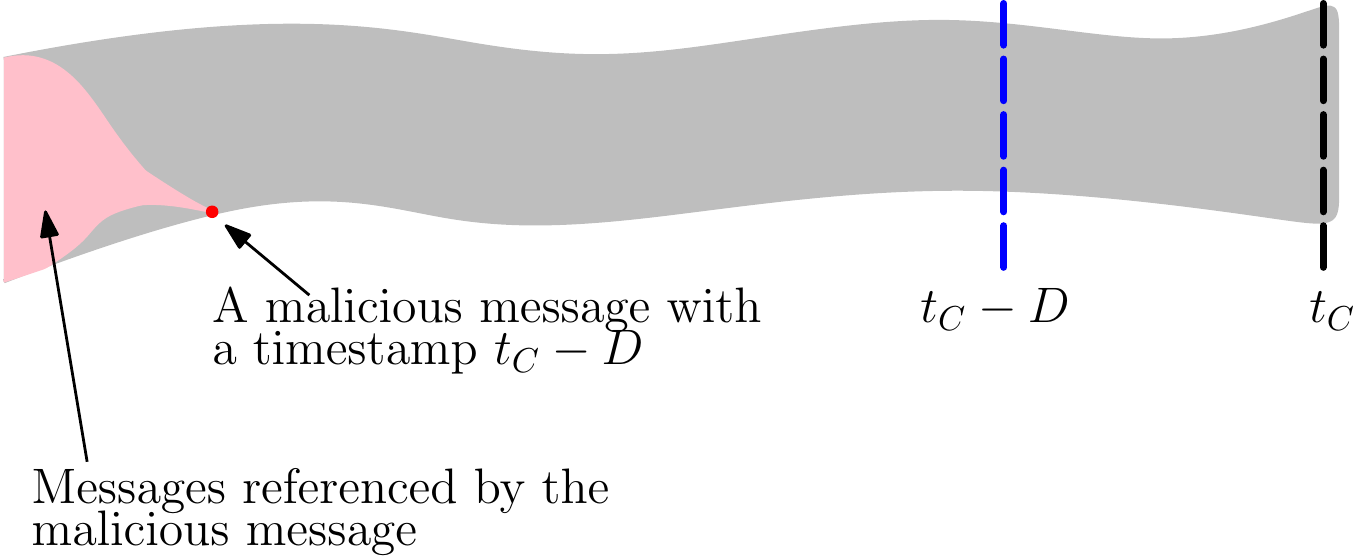}
  \caption{An example of a checkpoint manipulation in the absence of the random factor $X$. A malicious transaction (red) with a timestamp $t_{C} - D$ is issued ``deep'' in the DAG, even though honest transactions with these timestamps are placed near the blue dashed line. This placement of checkpoints could promote attacker's and diminish other user's reputation as only contributions from the pink ``cone'' would be used.   }\label{atta}
    \end{center}  
\end{figure}

The reputation changes, and it is possible that an attacker at one point in time had a high reputation but lost it in the meantime.  Then, an attacker tries to place checkpoint in the favorable (for him/her) part of the DAG. Without the random factor $X$, the timestamp of checkpoint would be predictable $t_{C} -\Delta$. An attacker could mine message with this timestamp and minimal hash. By placing it strategically in the DAG, an attacker could manipulate reputation obtained from summed over all of the messages approved by a checkpoint (points P\ref{ppp4} and \ref{pastcone}). This type of manipulation is presented in Fig. \ref{atta}.

Note that when the DAG is equipped with POG messages, then one of the POG messages can be used as a checkpoint.

Algorithm \ref{a2} describes the pseudocode for determining the committee with the checkpoint selection method.







\begin{algorithm}[t]
\DontPrintSemicolon
\Require{
    nodes\_reputation(time) $\rightarrow$ {\it descending ordered list of nodes' reputation at input time;}\\
    get\_random\_number()  $\rightarrow$ {\it global random number;}\\
    top\_nodes(n, rep\_list)  $\rightarrow$ {\it top n nodes in terms of reputation;}
    }
\Input{
    $t_C$: {\it time of committee selection;}\\
    $D$: {\it network delay;}\\
    n: {\it size of the committee;}\\
    }
\Output{
    committee: {\it list of nodes selected as committee members};
}
committee  $\leftarrow$ \{\}\;
\If{time = $t_C$} { 
    x $\leftarrow$ get\_random\_number()\;
    rep\_list $\leftarrow$ nodes\_reputation$(t_C - D -x)$\;
    committee $\leftarrow$ top\_nodes(n, rep\_list)\;
}
\textbf{return committee}
\caption{Committee selection scheduler for checkpoint selection method. }
\label{a2}
\end{algorithm}

\subsection{Maximal reputation
}

Maximal reputation from an interval approach is a combination of application message and checkpoint selection. The reputation value of the node $x$ is the maximal reputation calculated from all of the messages in the interval $[t_{C}-D -\Delta_A, t_{C}-D]$. This approach does not require the node $x$  to issue any special message. Note that for $\Delta_A=0$, it reduces to checkpoint selection without random factor. 

However, the method has higher computational complexity since it requires the computation of reputation for all messages issued in $[t_{C}-D -\Delta_A, t_{C}-D]$.

\section{Threat model and security}  \label{sec:secur}
The fact that the committee is only a subset of all the nodes may decrease the robustness against malicious actors. The security of the protocol depends on the size of the committee but also on the way the reputation is distributed among the nodes. 

\subsection{Zipf law}
Different protocols might define reputation and methods of gaining it differently. This affects the concentration of reputation and makes it impossible to perform an analysis in full generality. For this reason we propose to model the distribution of reputation using Zipf laws.

Zipf laws satisfy a universality phenomenon; they appear in numerous different fields of applications and have, in particular, also been utilized to model wealth in economic models \cite{wealth_pareto}. In this work we use a Zipf law to model the proportional reputation of $N$ nodes: the $n$th largest value $y(n)$ satisfies
\begin{equation}\label{eq:zipf}
    y(n)=C(s,N)^{-1} n^{-s},
\end{equation}
where $C(s,N)=\sum^N_{n=1} n^{-s}$, $N$ is the number of nodes, and $s$ is the Zipf parameter. A convenient way to observe a Zipf law is by plotting the data on a log-log graph, with the axes being log(rank order) and log(value). The data conforms to a Zipf law to the extent that the plot is linear, and the value of $s$ can be found using linear regression. 

\begin{figure}[ht]
\begin{center}   
  \includegraphics[width=0.7\textwidth]{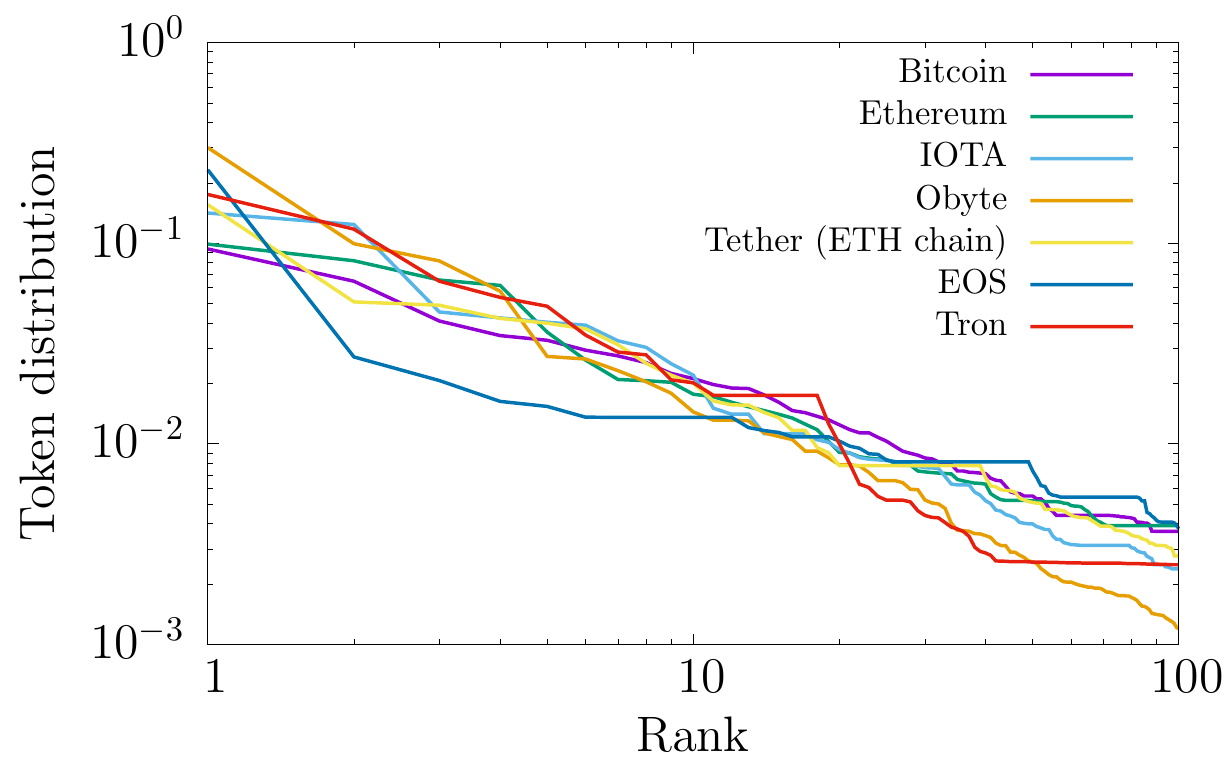}
  \end{center}  
  \vspace{-0.5cm} 
   \caption{The token distribution of selected cryptocurrency projects on a log-log scale (June 2020). }\label{zipf}
\end{figure}

In the Fig.\ \ref{zipf} we present the distribution of the richest accounts for a series of cryptocurrency projects for the top holders, which might be considered for the committee. We observe that most of them resemble Zipf laws. Table \ref{tab1} contains estimations of the corresponding coefficients of Zipf law. Note that for PoS-based protocols the reputation of a node can, to some extent, be approximated by the distribution of the tokens. Post-Coordicade IOTA uses a reputation system called \emph{mana} that shares some similarities with a delegated PoS, see \cite{coordicide}, and it is reasonable to assume that the future distribution of mana would be Zipf like.

\begin{table}
\begin{centering}
\begin{tabular}{|c|c|}
\hline
Name   & Zipf coefficient \\
\hline
\hline
Bitcoin & 0.7628    \\
\hline
Ethereum & 0.756786    \\
\hline
IOTA & 0.934275     \\
\hline
Obyte & 1.14361    \\
\hline
Tether (ETH chain) & 0.815054    \\
\hline
EOS & 0.536744    \\
\hline
Tron & 1.02043    \\
\hline

\end{tabular}
\vspace{0.5cm} \caption{Coefficients of Zipf distribution with the best fit to the token distribution of given cryptocurrency project. Method: linear regression on a log-log scale (June 2020).  }\label{tab1}
\end{centering} 
\end{table}

\subsection{Overtaking the committee
}
We assume that an adversary possesses $q\%$ of the total reputation and that it can freely distribute this reputation among arbitrary many different nodes. Honest nodes share $(1-q)\%$ of the reputation. We assume that the reputation among the honest nodes is distributed according to a Zipf law. 

Overtaking of the committee occurs when the attacker gets $t$ threshold committee seats; the exact value of $t$ might depend on the particular application of the committee. 

\begin{figure}
\begin{center}   
  \includegraphics[width=0.7\textwidth]{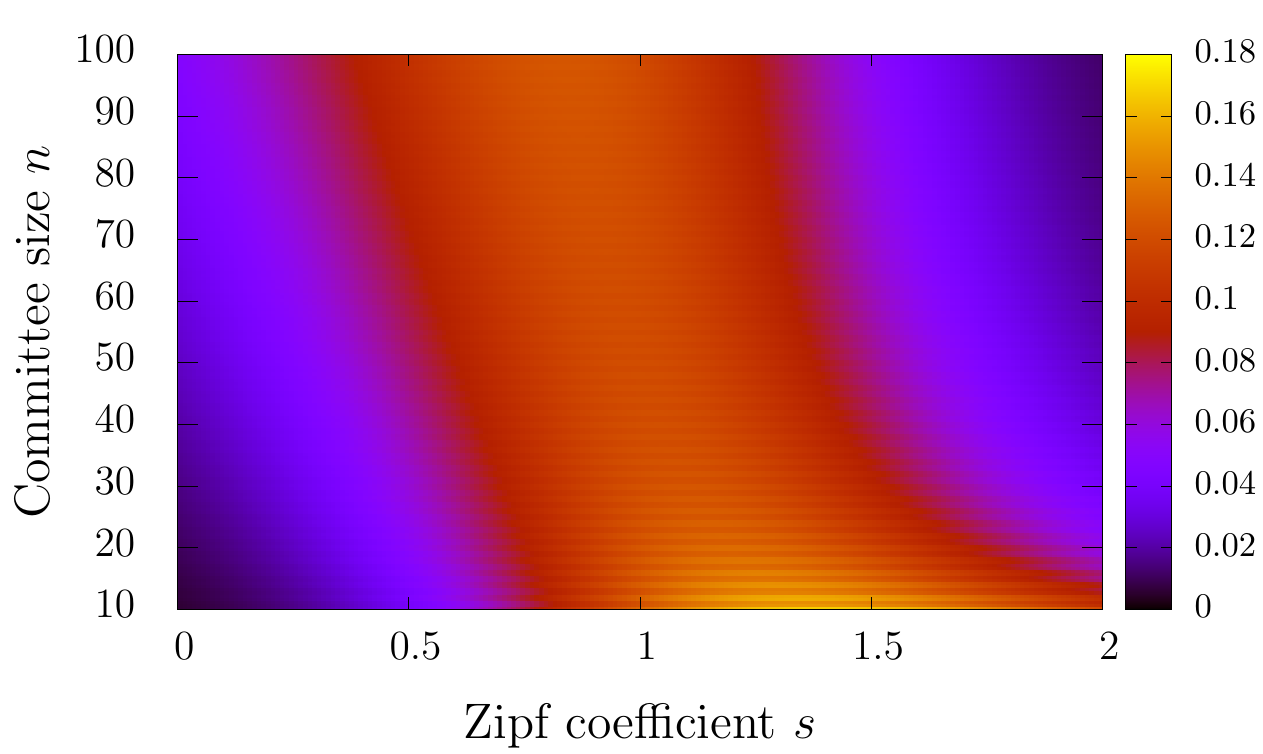}
  \end{center}  
   \caption{Minimal amount of reputation required to overtake the committee for different committee sizes and token distribution modeled with a Zipf distribution, threshold $t= \lfloor n/2 \rfloor + 1$, and number of total nodes $N=1000$. }\label{fig:q}
\end{figure}

The cheapest way for an attacker to obtain $t$ seats in the committee is to create $t$ nodes with $y(n-t+1)(1-q)$ reputation each.  The critical $q_c$ that allows to get $t$ seats in the committee is, therefore, given by 
\begin{equation}
    q_c= t \cdot y(n-t+1) (1-q_c).
\end{equation}
Equivalently,
\begin{equation}
    q_c= \frac{t\cdot y(n-t+1)}{1+t\cdot y(n-t+1)}.
\end{equation}
In Fig.\ \ref{fig:q} we present the critical value $q_c$ for reaching the majority in the committee.

\section{Applications} \label{sec:app}
In this section, we describe a few applications of the proposed committee selection protocols. 

\subsection{Decentralised random number generator}\label{sec:dRNG}

Committee selection protocols allow for the construction of a fully decentralized random number beacon embedded in the DAG ledger structure. An advantage of such an approach is that it relies only on the distributed ledger and does not require any interaction outside the ledger.

Publication of the random numbers would naturally take place in the ledger, where possible interruptions are publicly visible. Moreover, propagation of the random numbers would use the same infrastructure as the ledger.  

Gossiping among nodes in the network would decrease the committee nodes
communication overhead as they would not need to send randomness to each interested user. Similarly, if the proposed protocol requires a distributed key generation (DKG) phase (or any other setup phase), messages should be exchanged publicly in the ledger. Such an approach allows for the detection of malicious or malfunctioning committee members who did not participate in the DKG phase. After time $D$, a lack of corresponding messages can be verifiable proven using the ledger. 

This procedure increases robustness in the case of DKG phase failure. If such failure is detected, the committee selection is repeated until the DKG phase is successful. Any iteration of the selection process may not take into account the nodes that did not participate in the DKG phase previously. Moreover, if all communication is verifiable on the ledger, such nodes can be punished, e.g., by a loss of their reputation.

Note that if the dRNG protocol uses a $(t,n)$-threshold scheme, i.e.,  $n$ committee nodes publish their parts of the secret in the form of a beacon message and if $t$ or more beacon messages are published then the next random number can be revealed. The procedure of obtaining the random number from the beacon messages requires specific calculations, e.g., Lagrange interpolation. To save every node from performing those calculations, special nodes in the network can gather beacon messages and publish {\it collective beacon} messages which already contain the random number. The public can then verify these collective beacon messages against the collective public key.

\subsection{Smart contract oracles}

One main limitation of smart contracts is that they cannot access data outside of the ledger. So-called oracles try to address this problem by providing external data to smart contracts. When multiple parties engage in a smart contract, they have to determine an oracle. An obvious solution is for contracting parties to agree upon the oracle. However, this poses certain problems as oracles may be centralized points of failure and because the different parties have to find consensus on the choice of the oracle for each contract. Moreover, contractees must know in advance that the selected oracles are going to provide reliable services upon contract expiry. 

The committee selections proposed in Section \ref{sec:prot} can be used to improve the security and liveness of smart contracts. For instance, if oracles gain reputations that are recorded in the ledger, then oracles can be determined using the proposed methods. If the committee selection process occurs only upon the contract termination, then the majority, if not all of the selected oracles still provide services. 
A similar approach is adopted in blockchain-based Chainlink \cite{chainlink}. Our paper allows for using analogical methods in DAG-based DLTs.

This protocol would also be user friendly as contractees are not required to know classifications of oracles. Note that specialization is very likely to occur, and different oracles probably will deliver different types of data depending on the industry and requirements. For example, sports bets are settled with {\it sports reputation}, whereas events in a stock market may use {\it financial reputation}.

Further nodes can modify smart contracts themselves to modify the weight of oracles vote power or even make outcomes depend on the oracles' opinion distribution.

\subsection{Consensus and sharding}
The problem of consensus is much simpler in closed systems, where the number of participants is known and does not change. Unanimity algorithms that work in such an environment include \cite{HB,pBFT,mirBFT}. However, the closed nature of such protocols makes them not very relevant for decentralization. PoW is open for any user who is willing to solve the cryptographic puzzle; the network does not require any prior knowledge of the user. An intermediary step between entirely open and permissionless networks is a system where each user is allowed to set up a node and collect reputation, but only the most reliable nodes contribute to the consensus protocol. An example of such a protocol is EOS. In EOS delegated stake plays the role of the reputation. EOS blockchain database expands as a committee of $21$ validators with the highest DPoS produce blocks. Validators use a type of asynchronous Byzantine Fault Tolerance to reach consensus among themselves and propagate the new blocks to the rest of the network \cite{EOS}. 

\begin{figure}[ht]
\begin{center}
\begin{tikzpicture} [node distance=2.5cm,auto,>=latex'] 
\usetikzlibrary{positioning, fit, calc}   
\tikzset{block/.style={draw, ultra thick, text width=2.5cm ,minimum height=1.3cm, align=center},   
line/.style={-latex}     
}

\node[block] (a)  {Open, decentralized protocol};  
\node[block] at ([yshift=-3cm]$(a)$) (b) {Open protocol with a measure of: stake, honesty, importance, or reliability};   
\node[block]at ([yshift=-3cm]$(b)$) (c) {Closed system with a fixed number of participants};  

\draw[ultra thick, line] (a) edge node {Reputation gaining} (b);
\draw[ultra thick,line] (b) edge node {Committee selection} (c);

\end{tikzpicture}  

\end{center}
\caption{Process of finding fixed-size closed committee in open permissionless systems with reputation. } \label{open-closed}
\end{figure}
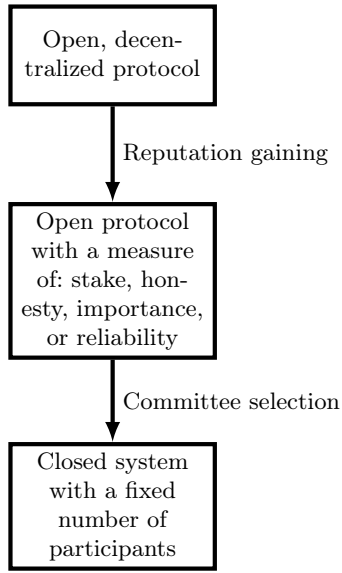

An illustration of the procedure of establishing consensus based on the fixed-size closed committee in open and permissionless systems is in Fig. \ref{open-closed}.

Most of the DLTs with committee based consensus use either pre-selected committees or a blockchain as the underlying database structure (EOS \cite{EOS}, TRON\cite{Tron}). Our committee selection protocols allow for using similar methods in a more general DAG structure (as long as DAG satisfies conditions P1-P5) . On the same note, proposed methods can also improve scaling through sharding solutions. It is straightforward to assigning validators to each shard based on their reputation. Similarly, as in the case of RepuChain \cite{repchain}, our solutions can assure that each shard has approximately the same total validator reputation.

\section{Conclusion and future research} \label{sec:dis} 

In this article, we proposed a committee selection protocol embedded in a DAG distributed ledger structure. We require the DAG to be equipped with an identity and reputation system. We further assumed that the ledger is immutable after some time, i.e.,  no transaction can be subtracted from the ledger; no transactions suggesting that it was issued a long time ago can be added to the ledger. These assumptions can be achieved by enforcement of approximately correct timestamps or POG transactions.

We further discussed methods of reading the reputation from the DAG. Methods include: (1) reputation summed over all of the messages approved by a given vertex in the DAG; (2) reputation summed over all of the messages with timestamps smaller than the timestamp of a given message.

Based on that we proposed and discussed the following methods of committee selection: application message, checkpoint selection, and maximal reputation method.

Furthermore, we analyzed the token distribution of a series of cryptocurrency projects, which turned out to follow Zipf distribution with a parameter $s \approx 1$. We used this fact to model reputation and analyzed the security of our proposal. Then, we focused on the applications of our committee selection protocols. We proposed an application to produce a fully decentralized random number beacon, which does not require any interaction outside of the ledger. Then, we showed how it could improve the oracle in smart contracts, and finally, we discussed possible advancements in consensus mechanism and sharding solutions.


Interesting research directions to further improve security and liveness of the proposed protocols might involve the use of backup committees. These solutions could be used when the primary committee is not fulfilling its duties. For instance, an obvious, although more centralized option, is a pre-selected committee controlled by the community or consortium of businesses interested in a reliable protocol. A different option is to select another reputation based committee from nodes with a reputation index  $\{n+1,...,2n\}$. However, the security of this solution is debatable, as it might be easier for an attacker to overtake it. Figure \ref{fig:q} shows that for $n=20$ and $t=11$ an attacker can overtake the committee with as little as $10\%$ of the reputation. The value of the secondary committee would be even lower.

Other improvements to the security of the committee include giving multiple seats to the top reputation nodes in the committee, e.g., nodes from $\{1,...,n/2\}$ would get double or triple identities in the committee. 
These adaptations increase the reputation requirement to overtake the committee. Other improvements of the liveness can include recovery mechanisms when the committee fails to deliver.

We hope that all mentioned improvements to the security and liveness will stimulate further research on this topic.

\bibliographystyle{splncs03}
\bibliography{sample-base}

\end{document}